**Deterministic and Scalable Coupling of Single 4H-SiC Spin Defects into Bullseye Cavities**

Tongyuan Bao[1,†], Qi Luo[1†], Ailun Yi[3,4†], Yingjie Li[1], Haibo Hu[1,2], Xin Ou[3,4*], Yu Zhou[1,5*], Qinghai Song[1,2,5,6*]

1. Ministry of Industry and Information Technology Key Lab of Micro-Nano Optoelectronic Information System, Guangdong Provincial Key Laboratory of Semiconductor Optoelectronic Materials and Intelligent Photonic Systems, Harbin Institute of Technology, Shenzhen 518055, P. R. China.
2. Pengcheng Laboratory, Shenzhen 518055, P. R. China.
3. State Key Laboratory of Functional Materials for Informatics, Shanghai Institute of Microsystem and Information Technology, Chinese Academy of Sciences, Shanghai 200050, China
4. The Center of Materials Science and Optoelectronics Engineering, University of Chinese Academy of Sciences, Beijing 100049, China.
5. Quantum Science Center of Guangdong-Hong Kong-Macao Greater Bay Area (Guangdong), Shenzhen 518045, China
6. Collaborative Innovation Center of Extreme Optics, Shanxi University, Taiyuan 030006, Shanxi, P. R. China.

† These authors contributed equally to this work.
*Corresponding authors: zhouyu2022@hit.edu.cn, ouxin@mail.sim.ac.cn or qinghai.song@hit.edu.cn

**Abstract:**
**Silicon carbide (SiC) has attracted significant attention as a promising quantum material due to its ability to host long-lived, optically addressable color centers with solid-state photonic interfaces. The CMOS compatibility of 4H-SiCOI (silicon-carbide-on-insulator) makes it an ideal platform for integrated quantum photonic devices and circuits. However, the deterministic integration of single spin defects into high-performance photonic cavities on this platform has remained a key challenge. In this work, we demonstrate the deterministic and scalable coupling of both ensemble (PL4) and single PL6 spin defects into monolithic bullseye cavities on the 4H-SiCOI platform. By tuning the cavity resonance, we achieve a 40-fold enhancement of the zero-phonon line (ZPL) intensity from ensemble PL4 defects, corresponding to a Purcell factor of approximately 5.0. For deterministically coupled single PL6 defects, we observe a threefold increase in the saturated photon count rate, confirm single-photon emission, and demonstrate coherent control of the spin state through optically detected magnetic resonance (ODMR), resonant excitation, and Rabi oscillations. These advancements establish a viable pathway for developing scalable, high-performance SiC-based quantum photonic circuits.**

**Main:**

Quantum photonic technologies, which aim to integrate single-photon sources[1] and spin-based quantum nodes[2–4] into scalable photonic circuits, drive the demand for efficient light-matter interfaces and robust photonic structures[5,6]. Achieving deterministic coupling of single quantum emitters into photonic cavities[7–11] is crucial for enabling applications in quantum communication and quantum networks[3,12–14]. Realizing such systems requires materials and platforms that combine excellent optical properties, CMOS compatibility, and scalable fabrication. Silicon carbide (SiC), with its wide transparent window, optical nonlinearities, and optically addressable spin defects, has emerged as a promising material platform for quantum photonics[15–20]. Spin defects such as divacancy and silicon-vacancy[21,22] in 4H-SiC offer long spin coherence times[23–25] and optically detectable magnetic resonance[26], making them highly suitable for spin-photon interfaces[27–29]. However, the lack of high-quality thin films on insulators has historically hindered the integration of 4H-SiC color centers into advanced photonic structures, as most implementations have relied on bulk carving techniques[8,30–32] that limit scalability and precision.

This challenge was recently addressed with the development of the 4H-silicon-carbide-on-insulator (4H-SiCOI) platform[33]. By leveraging wafer bonding and thinning techniques, high-quality 4H-SiC thin films have been demonstrated, enabling the fabrication of photonic devices with order-of-magnitude improvements in Q-factors. While the SiCOI platform has enabled significant progress in fabricating high-Q photonic devices, achieving deterministic coupling of single spin defects remains the next critical step for fully leveraging these advancements. For instance, with their unique circular Bragg grating structure, Bullseye cavities provide a powerful solution for efficient photon collection[34] and Purcell enhancement[35,36]. These cavities enable strong directional and enhanced ZPL emission, making them highly suitable for coupling single-photon emitters to far-field optical modes or coherent photon generation. However, integrating 4H-SiC spin defects into bullseye cavities has yet to be demonstrated, leaving a critical gap in realizing scalable SiC-based quantum photonic circuits.

Here, we demonstrate the deterministic integration of both ensemble PL4 and single PL6 spin defects into monolithic bullseye cavities on the 4H-SiCOI platform. By leveraging gas-tuning techniques, we precisely tune the cavity resonance, achieving a 40-fold enhancement in ZPL intensity for ensemble PL4 defects and observing a Purcell factor of 5.0. The cavity coupling enhances the saturation count rate of a single PL6 by a factor of 3 and reveals mutually orthogonal $E_x$ and $E_y$ coherent emission. Resonant and off-resonant optically detected magnetic resonance (ODMR) and Rabi oscillation measurements confirm the stability of the spin properties within the cavity environment. These results highlight the potential of 4H-SiCOI and bullseye cavities as scalable platforms for integrating quantum emitters into photonic circuits, paving the way for advanced SiC-based quantum photonic technologies.

**Scalable fabrication process for coupling single or ensemble SiC defects into cavities**

To achieve this, the first crucial step is to develop a robust and scalable fabrication process for reliably coupling spin defects to monolithic bullseye cavities on the 4H-SiCOI platform. As shown in Figure 1, the fabrication process begins with a 4H-SiC wafer, which features a 10 μm thick epitaxial layer and a thin SiO2 oxide layer. This oxide layer is fusion-bonded to an oxidized Si wafer and subsequently annealed. The resulting 400 μm thick SiC membrane is then thinned to a final thickness of 200 nm through a combination of grinding, mechanical polishing, and inductively coupled plasma (ICP) dry etching. For ensemble samples, carbon ions are implanted into the membrane to create SiC divacancy ensembles, followed by high-vacuum annealing. For single-center samples, gold markers are deposited on the surface to facilitate the coupling of individual divacancy spins, acting as reference points for the identification of single photoluminescent (PL) centers. The relative positions of these PL6 centers to the gold markers are determined using a home-built CCD (Charge Coupled Device) imaging system, with an accuracy on the order of tens of nanometers [37]. In the confocal scan (before and after coupling), the coupled single PL6 color centers are indicated with red arrows in the three figures. The bullseye cavities are designed with three free parameters: duty cycle, period length, and diameter of the central disk. By optimizing these three parameters, the resonant wavelength of the bullseye cavity can be designed to fall within the desired wavelength range (in our case, 1000 nm to 1100 nm). By keeping the duty cycle and period length constant, the resonant wavelength can be tuned by adjusting the diameter of the central disk, while maintaining a nearly constant resonance strength[38], as detailed in Supplementary Figure 1. We fabricated the devices by first depositing a 30 nm thick chromium (Cr) layer onto the SiCOI substrate. Electron beam lithography (EBL) is employed to pattern the bullseye cavity design onto the Cr layer, followed by the development of the resist and deposition of a Cr mask. This mask is then used to etch the photonic cavities into the SiC substrate. Finally, the Cr mask is removed, completing the fabrication of the bullseye cavities.

**Bullseye cavities coupled SiC PL4 spin ensembles**

With the devices successfully fabricated, we first characterized the cavities coupled to PL4 spin ensembles to validate our design and fabrication approach. Using carbon ion implantation, we successfully generate ensemble divacancy centers, evidenced by the photoluminescence (PL) spectra at 4K, as shown in Figure 1a (purple). The prominent zero-phonon line peak at 1080 nm belongs to PL4, indicating that this type of divacancy is preferentially formed during carbon ion implantation, consistent with previous findings [39,40]. The cavity mode of the bullseye structure can be directly measured from the PL spectrum of SiC spin ensembles. As depicted in Supplementary Figure 1, by varying the central disk diameters from 1.140 μm to 1.708 μm, the cavity mode can be tuned from $1024.86\pm0.03$ nm to $1122.66\pm0.04$ nm. More details about the design can be found in Supplementary Material S1 & S2. This enables coupling of the zero-phonon line (ZPL) and phonon sideband (PSB) of the PL4 ensembles to the cavity mode by adjusting the diameter of the central disk, as shown in the orange and green spectra in Figure 1a, respectively.

The cavity mode can be further tuned by introducing nitrogen gas into the cryostat chamber. For the PSB-coupled cavity, approximately 0.05 L of nitrogen gas (50 Pa) was injected in each step. As a result, the cavity mode was continuously tuned from $1120.31\pm0.02$ nm to $1124.03\pm0.03$ nm, exhibiting saturation behavior. For the ZPL-coupled cavity, 0.05 L of nitrogen gas (15 Pa) was injected in each step. While the shift in the cavity mode (marked by the dashed line) was less pronounced than in the PSB case, a significant enhancement in emission occurred when the ZPL mode intersected with the cavity mode. Points A-F on the map correspond to specific stages in the gas condensation process. At point C, where the cavity mode best matches the ZPL, the intensity was enhanced by 40 fold. The lifetime of the filtered ZPL photons was also measured at various stages and summarized in Figures 1b and 1c. As expected, the enhancement was more pronounced when the cavity mode approached the ZPL, resulting in a shorter lifetime. A comparison between Point F and Point C reveals that the lifetime decreased from $15.18\pm0.10$ ns to $13.18\pm0.03$ ns. The relationship between the Purcell factor and the lifetime decay is given by[41]

$$F = \frac{1}{\mathrm{DFT}}\left(\frac{\tau_{\mathrm{off}}}{\tau_{\mathrm{on}}} - 1\right),$$

where $\tau_{\mathrm{off}}$ and $\tau_{\mathrm{on}}$ are the lifetime corresponding to the resonance and off-resonance conditions of the ZPL, respectively, DWF denotes the Debye-Waller factor of the color center. Based on the uncoupled spectrum shown in Fig. 2(a), the DWF of PL4 divacancy is calculated to be 0.031, corresponding to a Purcell factor of 4.9.

**Bullseye cavities coupled single PL6 spin**

Building on the successful demonstration of coupling at the ensemble level, we advanced our investigation to the single-emitter regime by deterministically integrating individual PL6 spin defects—a critical step toward scalable quantum photonic devices. At 4K, a confocal scan map of the bullseye cavity, as depicted in Figure 3a, suggests that the single PL6 color center has been successfully integrated into the center of the fabricated bullseye cavity. The Second-order correlation measurements show a value of $g^2(0)$ well below 0.5, indicating the emission of a single-photon nature, as shown in Figure 3b. The PL spectrum measurement of the single spin exhibits a two-peak structure at 1041.95±0.01 nm and 1043.06 ± 0.02 nm in Figure 3c. We attribute this splitting of the excited states to the strain in the cavity[42]. The saturation behavior of the single PL6 center is measured both with and without the bullseye cavity. A threefold increase in saturation intensity ($44.74\pm0.91$ kcps to $91.04\pm0.60$ kcps) is observed in the bullseye cavity coupled single spin. Generating coherent photons is critical for most quantum photonic applications[6,43–45] and enables detailed study of the cavity-coupled single PL6's optical transitions. A tunable narrow linewidth diode laser in the set sketch graph in Figure 4a was used to excite the defect resonantly at 4K. The resonant excitation spectrum in Figure 4a also shows two peaks as the PL spectrum measurements in Figure 3c, corresponding to the Ex and Ey transitions of the excited state at 1042.57±0.01 nm and 1043.67±0.04 nm, with linewidths of 10.37±0.09 GHz and 9.06±0.32 GHz, respectively[42]. The polarization characteristics of Ex, Ey are further confirmed by polarization-dependent emission measurements, which are approximately orthogonal to each other. A detailed characterization of the spin and fluorescence properties of the single PL6 color center is presented in Supplementary S3.

To study the stability of the transition, the resonant spectrum is repeated by performing 12 consecutive scans under the same laser power as displayed in Figure 4d. To verify the spin properties of the single PL6 were well preserved in the cavity, the optically detected magnetic resonance (ODMR) spectra of the single PL6 center within the bullseye cavity are obtained under different excitation conditions: non-resonant laser excitation, resonant excitation of the Ex transition, and resonant excitation of the Ey transition as illustrated in Figure 5b. All ODMR spectra reveal two peaks at frequencies of 1351.71±0.18 MHz and 1358.56±0.12 MHz, which correspond to the spin states of the PL6 center. The contrast in the ODMR signal varies under different excitation conditions, with the highest contrast observed when the laser is resonant with the Ey transition. Coherent control of the PL6 spin is demonstrated through Rabi oscillations measured under off-resonant excitation, indicating successful coherent control of the single PL6 spin coupled to the bullseye cavity.

**Conclusion**

In conclusion, we have successfully demonstrated the deterministic integration and coherent control of spin defects within monolithic bullseye cavities fabricated on a scalable 4H-SiCOI platform. The successful demonstration of optically detected magnetic resonance and coherent Rabi oscillations for a cavity-coupled single spin confirms the robustness of this integrated system for quantum state manipulation. These advancements lay the groundwork for developing more sophisticated SiC-based quantum photonic devices with improved functionalities, paving the way for future applications in quantum networking, sensing, and computation.

**Methods**
**4H-SiCOI and micro-ring cavity fabrication.**
A 4-inch 4H-SiC wafer with an epitaxial layer, featuring N-doping concentrations of $1\times10^{18}$ cm$^{-3}$ (wafer) and $5\times10^{14}$ cm$^{-3}$ (epi-layer), was prepared via the standard RCA cleaning process. For activation, both the wafer and an oxidized silicon substrate were exposed to 100 W $O_2$ plasma for 30 seconds. A bond was formed between the two wafers at room temperature with an applied pressure of 3000 N. This structure then underwent a 6-hour annealing cycle at 800 °C to fortify the bond ahead of grinding.

A two-phase grinding method was employed. In the first phase, a 10-hour grinding session with 10 μm diamond slurry under 600 N of pressure thinned the SiC layer to approximately 30 μm. In the second phase, the SiC was further reduced to 10 μm with a 3 μm diamond slurry. Subsequently, chemical mechanical polishing (CMP) was used to eliminate surface damage from grinding. After dicing the SiCOI wafer into individual dies, each die was brought to its final 200 nm thickness by ICP dry etching (100 W RF, 1000 W ICP, 10 mTorr). Ensemble divacancy spins were introduced through the implantation of 30 keV carbon ions ($1.0\times10^{13}$ cm$^{-2}$ dose), followed by a 30-minute, 900°C high-vacuum anneal to repair the crystal lattice.

Photonic device fabrication commenced with the deposition of a 30 nm Cr layer on the spin-ensemble SiCOI membrane. Electron-beam lithography, along with an overlay technique, was utilized to pattern micro-ring cavities of differing diameters onto a ZEP 520A resist. After resist development, an ICP dry etch formed the Cr mask. Finally, using this Cr hard mask, the SiC membrane was etched with an ICP process (100 W RF, 1000 W ICP, 10 mTorr) to create the photonic devices.

**Device characterization and spin control of ensemble spins.**

The properties of the micro-ring cavities integrated quantum defects were investigated with a home-built confocal microscope system operating at 5 K inside a Montana Cryostation (CryoAdvance 50). The optical path included a 0.65 NA near-infrared objective (Olympus, LCPLN50XIR) and a PHOTEC superconducting nanowire single-photon detector. Depending on the experiment, different laser sources were employed: a 914 nm diode laser (MIL-III_914-300mW) was used for PL4 divacancy spin excitation, while a Toptica DL pro laser was utilized during resonant fluorescence scans. For measurements of fluorescence lifetime, a picosecond 940 nm laser (NKT Photonics, PIL1-094-40FC) was paired with a time-correlated single-photon counting (TCSPC) system (ID1000). A 50 μm copper wire served as the microwave delivery antenna, and a pulse blaster (Spincore, PBESR-PRO-500-PCI) ensured precise timing and synchronization of all sequences.

**Data availability**

Source data to generate figures and tables are available from the corresponding authors on reasonable request.

**Acknowledgments**

We acknowledge the support from the National Key R&D Program of China (Grant No. 2021YFA1400802, 2022YFA1404601, 2023YFB2806700), the Innovation Program for Quantum Science and Technology (No. 2024ZD0302100), the National Natural Science Foundation of China (Grant No. 12304568, 11934012, 62293520, 62293522, 62293521, 12074400 and 62205363), the GuangDong Basic and Applied Basic Research Foundation (Grant No. 2022A1515110382), Shenzhen Fundamental research project (Grant No. JCYJ20241202300152, J20230807094408018), Guangdong Provincial Quantum Science Strategic Initiative (GDZX2403004, GDZX2303001, GDZX2306002, GDZX2200001, GDZX2406002), Young Elite Scientists Sponsorship Program by CAST, New Cornerstone Science Foundation through the XPLORER PRIZE, Shanghai Science and Technology

Innovation Action Plan Program (Grant No. 22JC1403300), CAS Project for Young Scientists in Basic Research (Grant No. YSBR-69).

**Author contributions**
Q.S. and Y.Z. conceived the idea. A.Y. and X.O. prepared the SiC membrane. Q.L. and T.B. carried out the EBL lithography and Device fabrication. T.B. and Y.Z. built the setup and carried out the measurements. T.B., Y.Z., and Q.S. performed the simulations. Y.Z., T.B., and Q.S. wrote the manuscript. All authors contributed to analyzing the data and commenting on the manuscript.

**Competing interests**
The authors declare no competing interests.

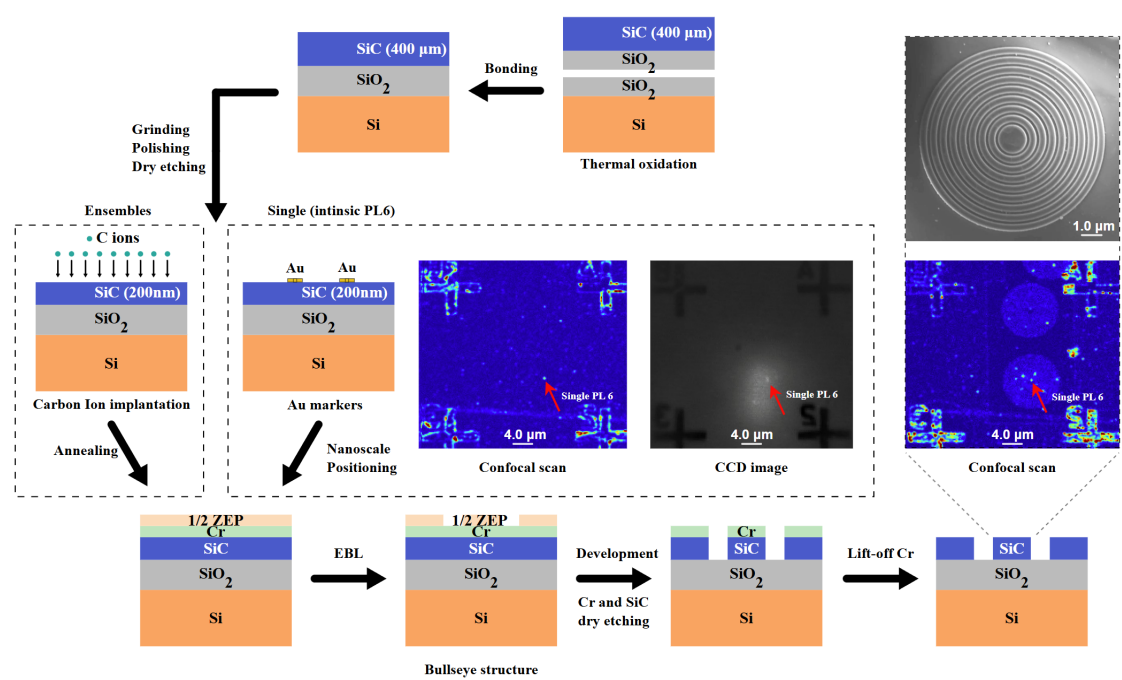

**Figure 1. The fabrication of 4H-SiCOI and the bullseye cavities on SiC membrane.** A 4H-SiC wafer with an epitaxial layer and a Si wafer were oxidized to form a thin layer of SiO2 on the surface. The two wafers were then fusion-bonded through annealing to enhance the bonding strength. The 400 μm thick SiC membrane was thinned to a target thickness of 200 nm through successive grinding, mechanical polishing, and ICP dry etching. One sample was used for injecting divacancy ensembles, which were introduced into the membrane via carbon ion implantation followed by high-vacuum annealing. Another sample was prepared by depositing gold markers on its surface to serve as reference points for identifying individual PL6 centers. The position of individual PL6 center relative to the gold markers was determined using CCD imaging of the center. The fabrication of bullseye cavities began with the deposition of a 30 nm Cr layer onto the SiCOI substrate containing the ensemble spins. Bullseye cavities were patterned onto an electron beam resist (ZEP 520A) using electron beam lithography (EBL). After the resist was developed, a Cr mask was created through ICP dry etching, which was used to etch the designed photonic devices into the SiC substrate. Finally, the Cr mask was lifted off, and the devices were fabricated.

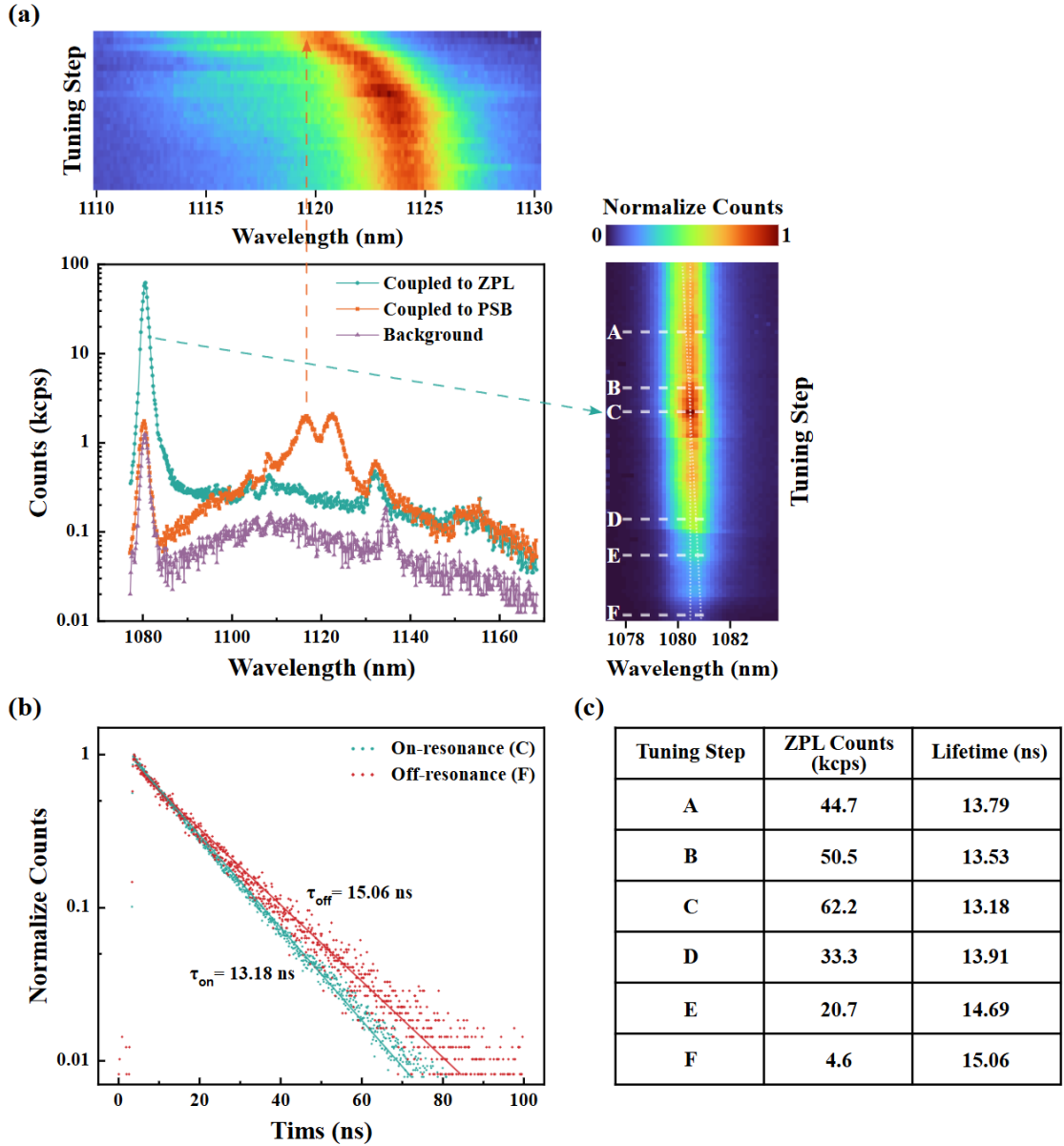

**Figure 2. Continuous cavity mode tunning via gas condensation. (a).** The PL spectrum were obtained from un-patterned SiC membrane and bullseyes of varying central disk diameters, which were coupled to the ZPL (1.668 μm) and PSB (1.260 μm) of the PL4 ensembles, respectively. The mode wavelength of the bullseye cavities was tuned using the gas condensation method. To the right and above the PL spectrum are intensity tuning maps corresponding to modes coupled to the ZPL and the PSB, respectively. For each cycle, approximately 0.05 L of nitrogen gas was introduced into the cryostat at pressures of 15 Pa and 50 Pa, respectively. The cavity mode undergoes red shifting while the ZPL wavelength remains constant. A significant enhancement in emission is observed when the two intersect on the intensity map. The high ZPL count after coupling to the mode results in negligible mode shifting, whereas the bullseye mode coupled to the PSB exhibits more pronounced shifts. Points A-F represent specific stages within the continuous gas condensation process. At the point C, the intensity of ZPL increase about ** times fold compare to the non cavity coupled ones.F **(b).** Lifetime measurements of the ZPL fitted with a single exponential function, indicating a reduction in lifetime from $15.18\pm0.09$ ns off-resonance to $13.18\pm0.03$ ns on-resonance. **(c)**. Summary of lifetimes at points A-F, demonstrating that the closer the cavity mode is to the ZPL, the more pronounced the reduction in lifetime.

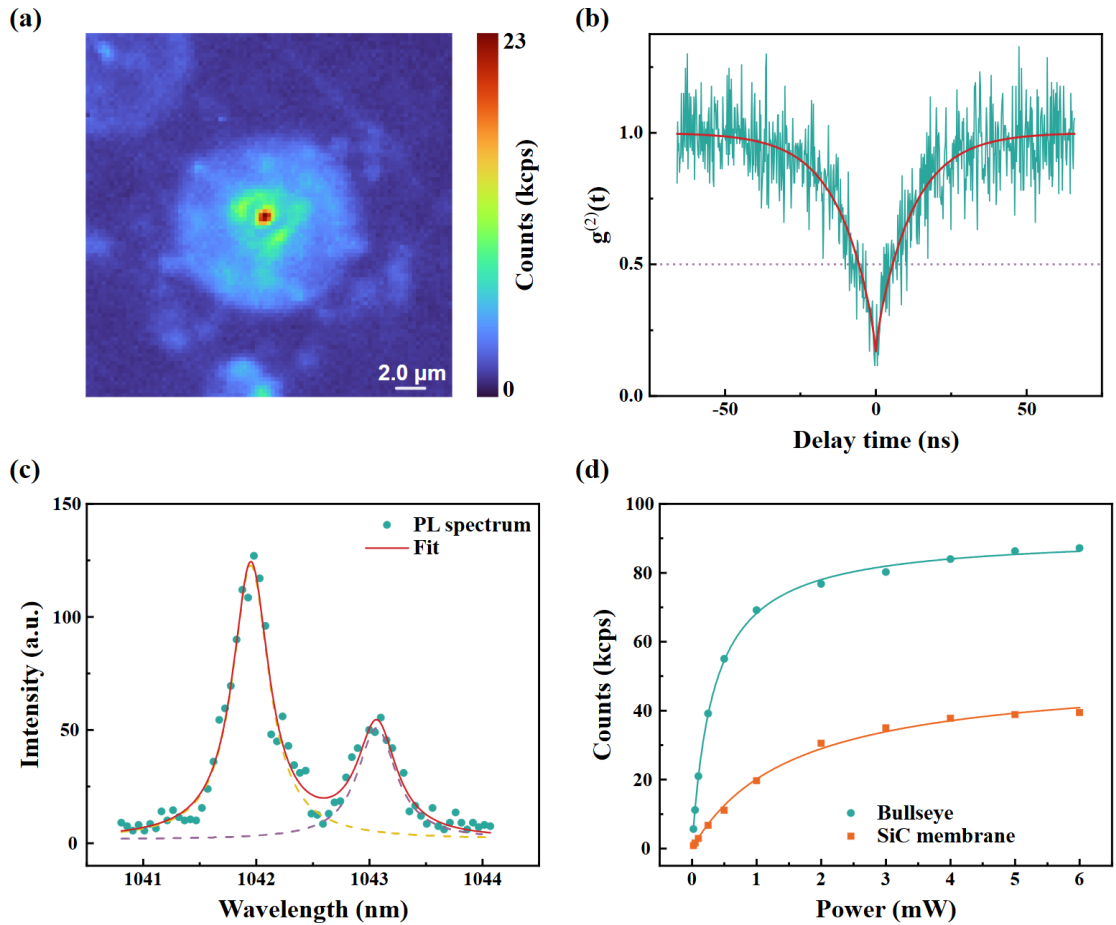

**Figure 3. Optical characterization of the single PL6 within the bullseye cavity at 4K. (a)**. The confocal scan map of the bullseye cavity at 4K. The map is shown rotated by 45° relative to the original CCD image for optimal viewing. **(b)**. The second-order correlation measurement of the PL6 within the bullseye cavity, with the excitation laser power of 100 μW, $g^2(0)$ is well below 0.5, indicating a signature of single photon emission. **(c)**. The ZPL of the PL6 within the bullseye cavity exhibits two distinct peaks at $1041.95\pm0.01$ nm and $1043.06\pm0.02$ nm, indicating that significant strain leads to the splitting of the excited-state energy levels into two branches. **(d)**. Saturation curves of single PL6 centers within and without a bullseye cavity. The counts are the background-corrected experimental data. The solid line fits with a function of $I(P)=I_s \cdot P/(P + P_s)$, where P and I(P) are the power of the excitation laser and the corresponding count rate, respectively, with $I_s$ and $P_s$ being the saturated intensity and saturated exciting power, respectively. The saturation intensity is increased by a factor of up to three.

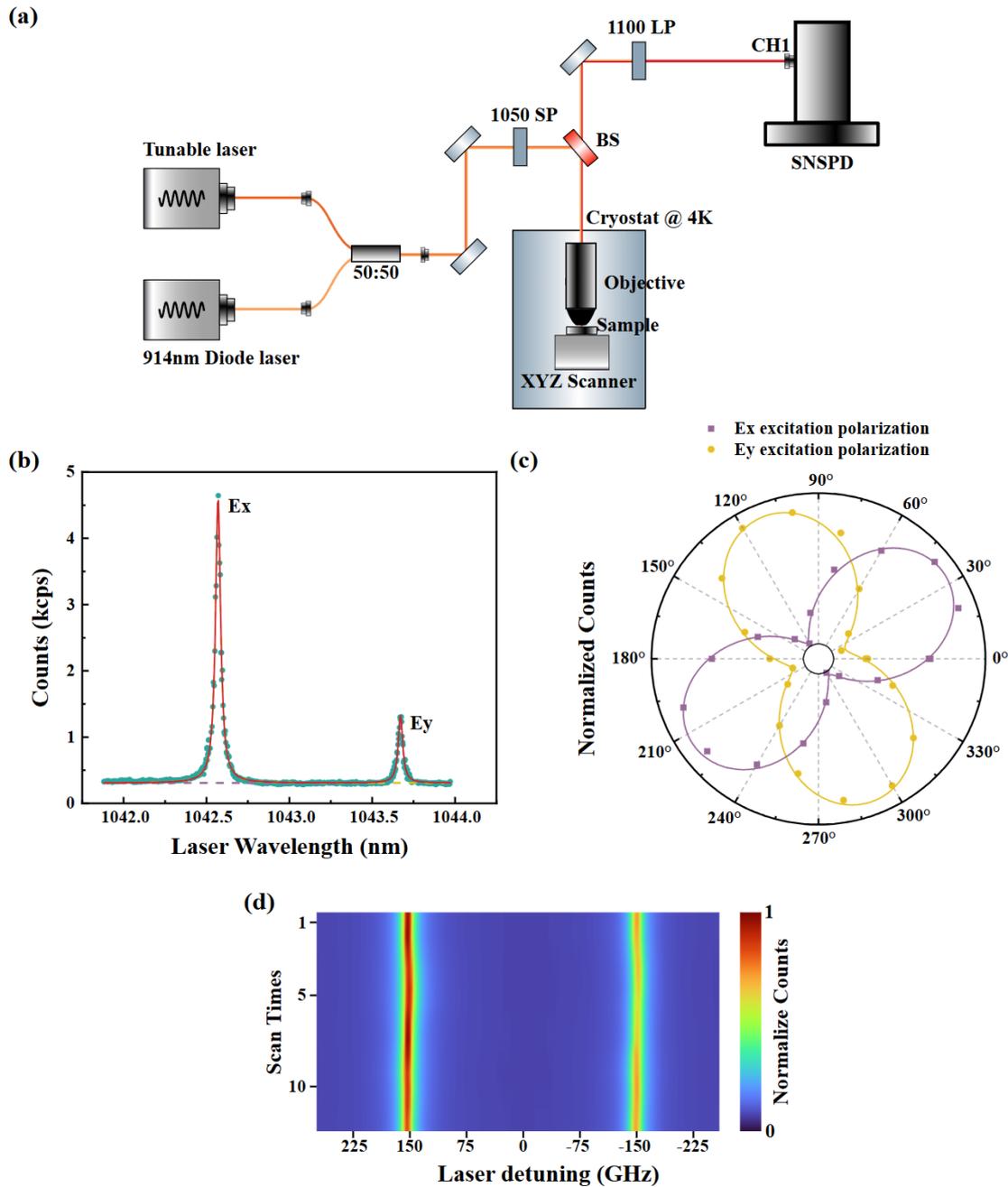

**Figure 4. Resonant excitation properties for the single PL6 within the bullseye cavity. (a).** Optical setups used for photoluminescence excitation. A tunable laser and a off-resonant 914 nm laser were combined via a 50:50 beam splitter to excite PL6, where the 914 nm laser was employed to restore the charge state of PL6. A custom-built 4K microscopy system collects the fluorescence from PSB of the PL6 spin. **(b).** The PLE spectrum of the single PL6 aligns well with its PL spectrum, with the off-resonant laser excitation at 1.3 μW and the resonant laser excitation at 0.5 μW, exhibiting two peaks located at 1042.57±0.01 nm and 1043.67±0.04 nm, corresponding to the Ex and Ey transition branches of the excited state energy levels, respectively. The linewidths of the Ex and Ey transitions are measured to be 10.37±0.09 GHz and 9.06±0.32 GHz, respectively. A Lorentzian fit to the raw data is displayed in red. **(c).** The excitation polarization characteristics of the Ex and Ey transitions in PL6. By tuning the laser to resonance with the transitions and continuously varying the polarization angle of the excitation laser, the corresponding photon counts were recorded. The solid lines represent the results of sinusoidal fitting. The excitation polarization angles of the Ex and Ey transitions are approximately orthogonal. **(d)**. Under the same laser power, 12 consecutive PLE scans spectra were performed on the PL6, demonstrating the stability of the Ex and Ey transitions.

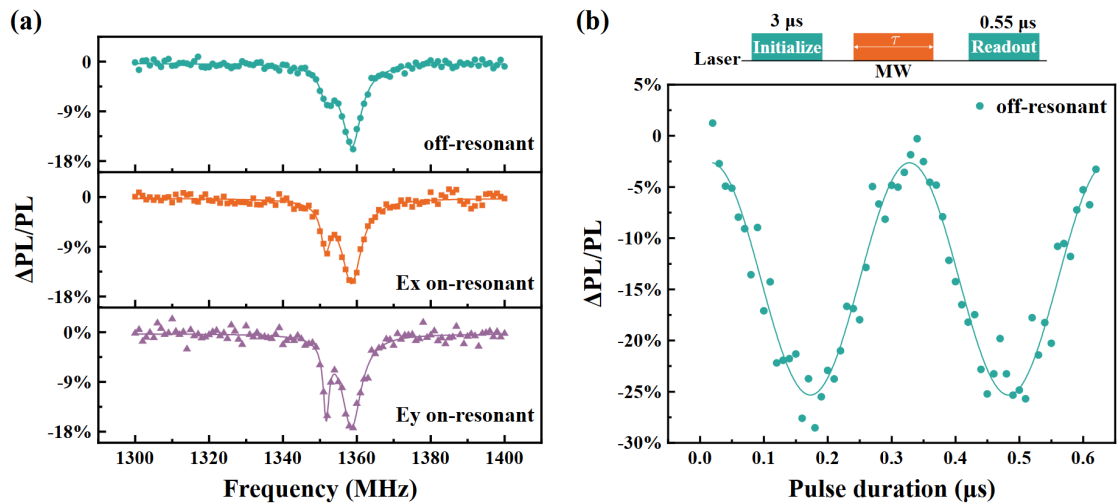

**Figure 5. ODMR and coherent control of the single PL6 spin in the bullseye cavity. (a)**. The three ODMR measurements were obtained under the following conditions: (i) non-resonant laser excitation, (ii) resonant laser excitation of the Ex transition, and (iii) resonant laser excitation of the Ey transition. All spectra reveal two peaks at 1351.71±0.18 MHz and 1358.56±0.12 MHz. From Lorentzian fitting, the ODMR contrast is extracted to be 5.1%± 0.4% and 14.7%±0.3%, 7.2%±0.7% and 15.1%±0.4%, 12.8%±1.2% and 16.7%±0.6%, respectively. **(b)**. Rabi oscillations of the PL6 divacancy spin under 100 µW off-resonant laser excitation, demonstrating coherent control of PL6 spin coupled to the bulleye cavity. The laser and microwave pulse sequences utilized for measuring the Rabi oscillations are illustrated.